\documentclass[a4paper]{article}
\usepackage{multirow}
\usepackage{booktabs}
\usepackage{INTERSPEECH2021}
\usepackage{stfloats}
\usepackage{color}
\usepackage{siunitx}
\usepackage[english]{babel}
\usepackage{hyperref}
\usepackage{cite}

\title{INTERSPEECH 2021 ConferencingSpeech Challenge:
Towards Far-field Multi-Channel Speech Enhancement for Video Conferencing}
\name{Wei Rao$^{1}$, Yihui Fu$^{2}$, Yanxin Hu$^{2}$, Xin Xu$^{3}$, Yvkai Jv$^{2}$, Jiangyu Han$^{1}$, Zhongjie Jiang$^{1}$, Lei Xie$^{2}$, Yannan Wang$^{1}$, Shinji Watanabe$^{4,5}$, Zheng-Hua Tan$^{6}$, Hui Bu$^{3}$, Tao Yu$^{7}$, Shidong Shang$^{1}$}

\address{$^{1}$Tencent Ethereal Audio Lab, China, 
            $^{2}$Northwestern Polytechnical University, China, \\
            $^{3}$Beijing Shell Shell Technology Co., LTD., China,
            $^{4}$Carnegie Mellon University, USA,
            $^{5}$Johns \\ Hopkins University, USA,
            $^{6}$Aalborg University, Denmark, 
            $^{7}$Tencent Ethereal Audio Lab, USA} 
             
\email{ellenwrao@tencent.com}

\begin{document}

\maketitle
\begin{abstract}
The ConferencingSpeech 2021 challenge is proposed to stimulate research on far-field multi-channel speech enhancement for video conferencing. The challenge consists of two separate tasks: 1) Task 1 is multi-channel speech enhancement with single microphone array and focusing on practical application with real-time requirement and 2) Task 2 is multi-channel speech enhancement with multiple distributed microphone arrays, which is a non-real-time track and does not have any constraints so that participants could explore any algorithms to obtain high speech quality. Targeting the real video conferencing room application, the challenge database was recorded from real speakers and all recording facilities were located by following the real setup of conferencing room. In this challenge, we open-sourced the list of open source clean speech and noise datasets, simulation scripts, and a baseline system for participants to develop their own system. The final ranking of the challenge will be decided by the subjective evaluation which is performed using Absolute Category Ratings (ACR) to estimate Mean Opinion Score (MOS), speech MOS (S-MOS), and noise MOS (N-MOS). This paper describes the challenge, tasks, datasets, and subjective evaluation. The baseline system which is a complex ratio mask based neural network and its experimental results are also presented.
\end{abstract}
\noindent\textbf{Index Terms}: ConferencingSpeech challenge, multi-channel speech enhancement,  multiple distributed microphone arrays, casual system, subjective evaluation.

\section{Introduction}

In recent years, video conferencing becomes increasingly important. It helps us to seamlessly connect with people of our choice anytime anywhere in the world and break barriers of distance among people. However, during video conference, the speech quality will be significantly affected by background noise, reverberation, the number of recording microphones, the layout of microphone array, the acoustic and circuit design of microphone arrays, interference speakers, and so on. Effective speech enhancement plays an important role in the video conferencing system. Although the performance of speech enhancement has been improved dramatically in the past several decades \cite{loizou2013speech,Gerkmann2015phase,Gannot2017Multi,michelsanti2020overview,xu2017weighted,weninger2015speech,chen2015speech,michelsanti2017conditional,kolbaek2016speech,chai2017gaussian,hao2019attention,hu2020dccrn}, there are still a set of challenging problems that should be further addressed in the far-field complex meeting room environments.

ConferencingSpeech 2021 challenge is proposed to stimulate research on processing the far-field speech recorded from microphone arrays in video conferencing rooms and has the following features: 1) Focusing on the far-field multi-channel speech enhancement problem using multiple distributed microphone arrays in the real meeting room scenario; 2) Exploring real-time multi-channel speech enhancement methods to achieve superior perceptual quality and intelligibility of enhanced speech with low latency and no future ``frame'' information; 3) Targeting the real video conferencing room application, in which the challenge database is recorded from real speakers in real conference rooms. Multiple microphone arrays with 3 different geometric topologies are allocated in each recording room. The number of speakers and the distances between speakers and microphone arrays vary particularly according to the sizes of meeting rooms. Twelve different sizes and decorated materials of rooms with the presence of common meeting room noises are used for recording, which makes the reverberation and noises as the dominating factors affecting the speech quality; 4) Focusing on the development of algorithms, the challenge requires the \textit{close} training condition. In other words, only provided list of open source clean speech datasets and noise dataset could be used for training; 5) The final ranking of the challenge will be decided by subjective evaluation. The subjective evaluation will be performed using Absolute Category Ratings (ACR) to estimate Mean Opinion Score (MOS), speech MOS (S-MOS), and noise MOS (N-MOS) by the subjective evaluation platform.

\subsection{Related Works}

Some corpora were released to promote the research on the far-field scenario. Earlier meeting corpora include ICSI \cite{janin2003icsi}, AMI \cite{carletta2005ami}, CHIL \cite{mostefa2007chil} and so on. The VOiCES corpus \cite{richey2018voices} is an open-sourced corpus focusing on distant speech under real room conditions. But the recordings were collected from distant microphones, not from microphone arrays. The LibriCSS \cite{chen2020continuous} corpus was proposed for speech separation task and consisted of multi-channel audio recordings recorded in the real room instead of being generated by simulation. However, the utterances were taken from LibriSpeech and played back from a loud-speaker placed in the room, not from real speakers. The CHiME-5 database \cite{barker2018fifth} simulated a dinner party scenario and collected distant multi-microphone speech recordings in everyday home environments. The DiPCo corpus \cite{van2019dipco} also imitated the dinner party scenario and collected the recordings by a single-channel close-talk microphone and five far-field 7-microphone array devices positioned at different locations in the recording room. 

Our ConferencingSpeech Challenge dataset is specially designed for the real video conferencing room and recorded from real speakers by multiple different types of microphone arrays. 12 meeting rooms with different sizes and decorative materials were used for recording. In addition, different from the recent challenges DNS \cite{reddy2020interspeech,reddy2020icassp} focusing on single-channel, CHiME-6 \cite{watanabe2020chime} focusing on multi-channel automatic speech recognition, and FFSVC2020 \cite{Qin2020ffsvc} focusing on multi-channel speaker verification, the ConferencingSpeech 2021 Challenge is proposed to explore far-field multi-channel speech enhancement methods to achieve superior perceptual quality and intelligibility of enhanced speech.

\section{Tasks and Rules}

The challenge consists of two tasks. There is no limitation on the system architecture, models, and training techniques. 

\textbf{Task1: Multi-channel speech enhancement with single microphone array}. This task focuses on processing the speech from single linear microphone array with non-uniform distributed microphones and considering practical application with real-time requirement. No future ``frame'' information could be used in Task 1. Frame length should be less than or equal to 40ms. The real time factor of algorithms must be less than or equal to one on the single thread of an Intel Core i5 machine clocked at 2.4GHz or equivalent processors. The real time factor $F_{rt}$ is formulated as follows:
\begin{equation} \label{eq:rtf}
F_{rt} = \frac{T_p}{T_t}
\end{equation}
where $T_p$ is the processing time of given test clip; $T_t$ is the length of test audio. 

\textbf{Task2: Multi-channel speech enhancement with multiple distributed microphone arrays}. This task focuses on processing the speech from multiple distributed microphone arrays. There are five microphone arrays from three different geometric typologies. All speech signals from these five microphone arrays are synchronized. This task is a non-real-time track and does not have any constraints so that participants could explore any algorithms to obtain high speech quality.

\begin{figure*}[htb]
  \centering
  \centerline{\includegraphics[width=14cm]{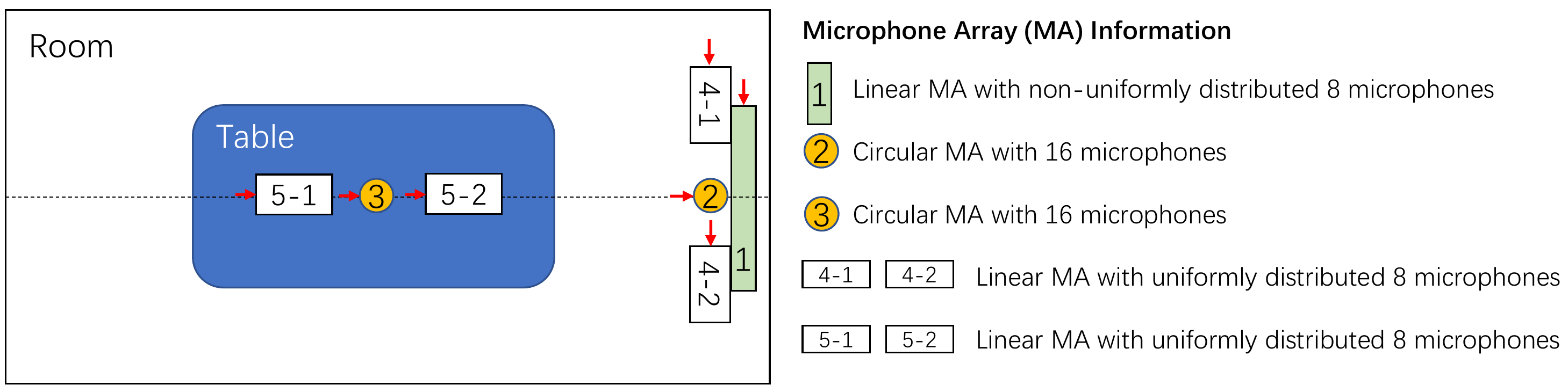}}
  \vspace{-5pt}  
\caption{The setup of microphone arrays in the meeting rooms. The red arrow points at the position of first microphone of each MA.}
  \vspace{-5pt}  
\label{fig:Setup_MA}
\end{figure*}

\begin{figure}[!htb]
	\begin{minipage}[b]{1.0\linewidth}
		\centering
		\centerline{\includegraphics[width=4cm]{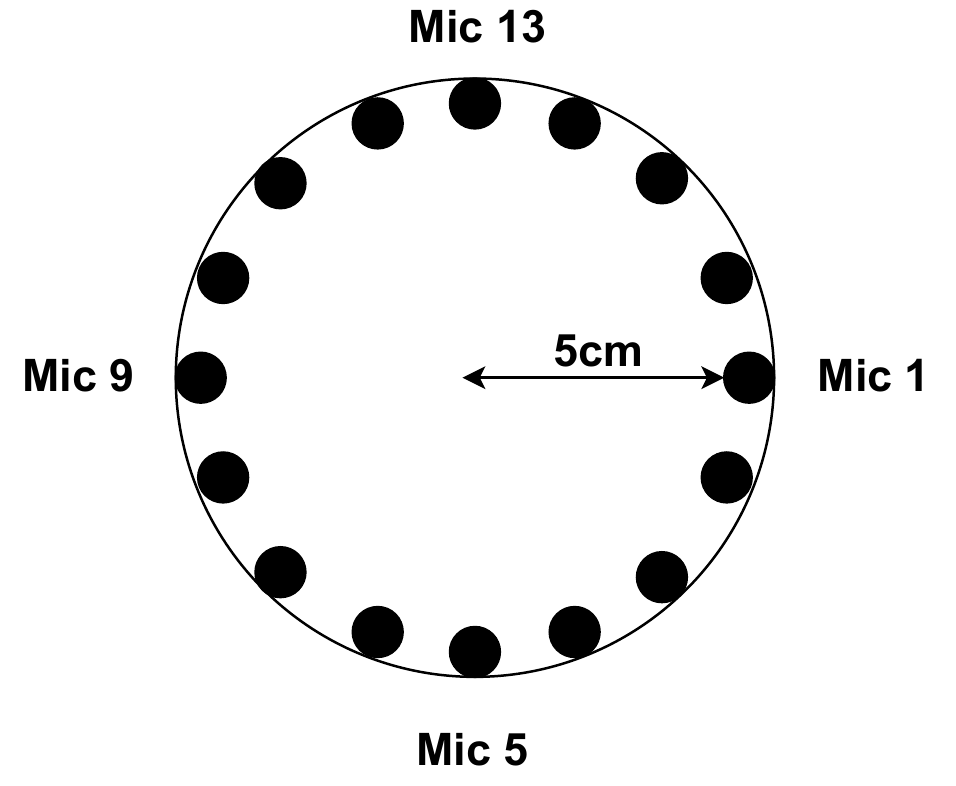}}
		\centerline{(a) Circular microphone array}\medskip
	\end{minipage}
	\begin{minipage}[b]{1.0\linewidth}
		\centering
		\centerline{\includegraphics[width=4cm]{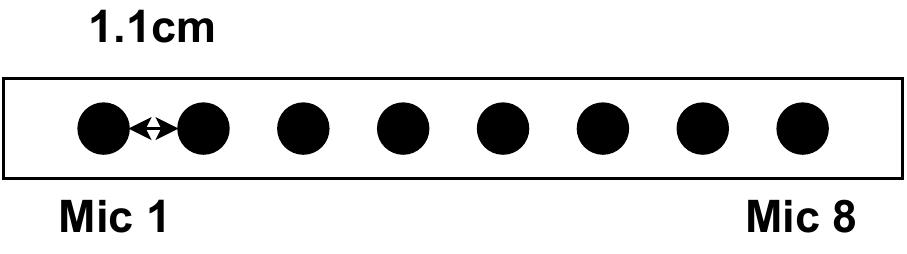}}
		\centerline{(b) Linear microphone array with uniformly distributed 8 microphones}\medskip
	\end{minipage}
	\hfill
	\begin{minipage}[b]{1.0\linewidth}
		\centering
		\centerline{\includegraphics[width=7.5cm]{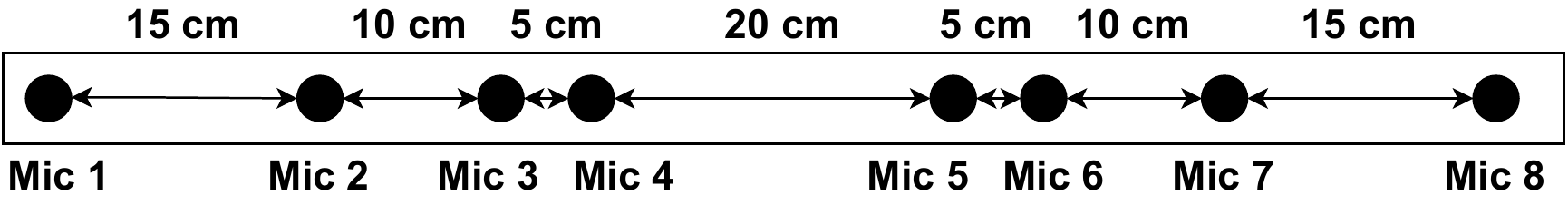}}
		\centerline{(c) Linear microphone array with non-uniformly distributed 8 microphones}\medskip
	\end{minipage}
	\caption{Configuration of three types of microphone arrays.}
	\label{fig:threearrays}
    \end{figure}

\section{Data Description}\label{Sec:data}

\subsection{ConferencingSpeech 2021 Challenge Database} \label{Sec:data-challenge}

Aiming at the real video conferencing room scenario, the ConferencingSpeech 2021 Challenge Database was recorded with real speakers and all recording facilities are located by following the real setup of video conferencing room. To simulate most of video conferencing room scenarios, 12 rooms with different sizes and decorative materials were used for recording. The decorative materials of rooms could be categorized into four type: without glass wall, with 1 glass walls, with 2 glass walls, and with 3 glass walls. The database contains two languages: English and Chinese. The sampling rate of recordings is 16kHz.

Multiple microphone arrays from three different geometry topologies were distributed in the recording rooms to investigate the impact of layout, acoustic and circuit design of microphone arrays on the speech quality. Figure \ref{fig:Setup_MA} shows the recording setup of ConferencingSpeech 2021 Challenge database. The recording devices included 5 microphone arrays from 3 different geometric topologies. The red arrow in Figure \ref{fig:Setup_MA} points to the first channel of microphone arrays. The microphone arrays in each meeting room followed the allocation of Figure \ref{fig:Setup_MA}, but the distances among microphone arrays (MAs) varied according to the sizes of meeting room. All recordings from these 5 MAs were synchronized. The information of MAs in Figure \ref{fig:Setup_MA} is summarized as follows:
\begin{itemize}
\item No.1 is a linear MA with non-uniformly distributed 8 microphones. The interval among microphones are illustrated in Figure \ref{fig:threearrays}(c).
\item No.2 and No.3 are circular MA with 16 microphones. The radius of circular MA is 5cm as shown in Figure \ref{fig:threearrays}(a). 
\item No.4 and No.5 are linear MA. Each MA is composed of two small linear MA with uniformly distributed 8 microphones. The interval among microphones in the small linear MA is 1.1cm as shown in Figure \ref{fig:threearrays}(b). 
\end{itemize}

The recording conditions of this database could be categorized into the following two parts:
\begin{itemize}
\item Semi-real recording: the data was recorded in the real meeting room scenario, but was not in the real meeting. The speech and noise data were separately recorded in all 12 quiet meeting rooms. Both were collected from playback and real speakers. Then, these speech and noise data were used for simulation. It contained two language: English and Chinese. The benefit of semi-real recording is that more variations of meeting scenarios could be considered. 
\item Real recording: the data was recorded during the real meeting in the real meeting room. All utterances were recorded from real speakers under real noise conditions. The recording language was Chinese.
\end{itemize}

\subsection{Training Set} \label{Sec:data-trn}
To focus on the development of algorithms, we designed the challenge with the \textit{close} training condition. In other words, only the provided list of open source clean speech datasets and noise dataset could be used for training.

\subsubsection{Clean Speech}
Clean training speech set signals were chosen from four open source speech databases: AISHELL-1~\cite{aishell_2017}, AISHELL-3~\cite{AISHELL-3_2020}, VCTK~\cite{christophe2016cstr}, and Librispeech (train-clean-360)~\cite{panayotov2015librispeech}. The speech utterances with SNR higher than 15 dB were selected for training. The total duration of clean training speech is around 550 hours. 

\subsubsection{Noise Set}
The Noise set is composed of two parts. Part I was selected from MUSAN~\cite{musan2015} and Audioset\footnote{\url{https://research.google.com/audioset/}}. The total duration is around 120 hours. Part II is the real meeting room noises recorded by high fidelity devices. The total number of clips is 98. 

\subsubsection{Room Impulse Responses (RIR)}

We used an image method to simulate RIRs for three microphone arrays introduced in Figure \ref{fig:threearrays}. The room size ranged from 3$\times$3$\times$3 $m^{3}$ to 8$\times$8$\times$3 $m^{3}$, containing more than 2500 rooms. The microphone array was randomly placed in the room with height ranging from 1.0 to 1.5 $m$. The sound source, including speech and noise, came from any possible position in the room with height ranging from 1.2 to 1.9 $m$. The angle between two sources was wider than \ang{20}. The distance between sound source and microphone array were ranged from 0.5 to 5.0 $m$. The total number of RIRs was more than 10,000 for each microphone array.

\subsection{Development set}

The development set was categorized into three parts: Simulation clips, Semi-real recordings, and Real recordings. Semi-real and real recordings are selected from the recordings in the ConferencingSpeech 2021 Challenge Database. 

\subsubsection{Simulation clips}

The simulation set was provided for participants to develop the systems and estimate the objective scores, which contains two sets: (1) a single MA set and (2) a multiple MA set. 

For the single MA set, we simulated 1,588 clips for three types of MA. The details can be found in Section~\ref{Sec:data-challenge}. Similar to the single MA set, multiple MA set also consistes of simulation clips from these three MAs. The only difference is that these three MAs are assumed in the same room during simulation.

1,624 clean speech selected from AISHELL-1, AISHELL-3, and VCTK and 800 noise clips selected from MUSAN were used for the simulation of both sets. The simulated SNR ranged from 0 to 30 dB and the duration of clips was 6 seconds.

\subsubsection{Semi-real recordings}

As mentioned in Section~\ref{Sec:data-trn}, the speech sources could be divided into playback and real speakers. The Semi-real recordings consisted of 2.35 hours of playback English speech segments and 2.31 hours of real speaker's Chinese speech segments. Each audio clip contained multiple channel information. All five MAs' recordings were provided for participants to develop their systems.

\subsubsection{Real recordings}

More than 200 real recording clips were provided, which are from 12 real speakers and their ages range from 18 to 50 years old. Similar to semi-real recordings, each audio clip contained multiple channel information and all five MAs' recordings were provided.

\subsection{Evaluation set}

The evaluation set consists of two task sets. We selected Audio clips from other 9 rooms' recordings of ConferencingSpeech 2021 Challenge Database which were unseen in development test set. No. 1 MA in Figure~\ref{fig:Setup_MA} was selected for Task 1. The recordings from five MAs in Figure~\ref{fig:Setup_MA} were provided for Task 2.  

Different from the development set, the evaluation set only contains the semi-real and real recordings. The evaluation set for each task is composed of three parts: semi-real recordings from playback, semi-real recordings from real speakers, and real meeting recordings. Specifically, in each task, 135 semi-real recordings from playback and real speakers were selected for each MA, respectively. And 165 real recordings were selected for each MA. 

\section{Subjective Evaluation}
The performance of each participated team was decided by subjective evaluation. Inspired from ITU-T P.835, the subjective evaluation was performed according to overall quality rating, speech signal rating, and background noise rating, for which we used Absolute Category Ratings (ACR) to estimate global Mean Opinion Score (MOS), Speech MOS (S-MOS), and Noise MOS (N-MOS), respectively. The details of evaluation metrics are as follows:
\begin{itemize}
\item MOS: Determination of subjective global MOS. The rater will select the category which best describes the overall quality of heard sample for the purpose of everyday speech communication. The categories of overall speech sample are 5-Excellent / 4-Good / 3-Fair / 2-Poor / 1-Bad. 
\item S-MOS: Determination of subjective speech MOS. The rater will only attend to the quality of speech signal. The categories of speech signal in this sample are 5-Not Distorted / 4-Slightly Distorted / 3-Somewhat Distorted / 2-Fairly Distorted / 1-Very Distorted.
\item N-MOS: Determination of subjective noise MOS. The rater will only attend to the background. The categories of background in this sample are 5-Not Noticeable / 4-Slightly Noticeable / 3- Noticeable But Not Intrusive / 2-Somewhat Intrusive / 1-Very Intrusive.
\item dMOS/dS-MOS/dN-MOS: Difference between the MOS/S-MOS/N-MOS after enhancement and MOS/S-MOS/N-MOS of the noisy evaluation set before enhancement.
\item CI: Confidence Interval of MOS score.
\end{itemize}

Each rater would determine MOS, S-MOS, and N-MOS scores for each evaluation audio file. And each evaluation audio file would be rated by more than 20 qualified raters. Two rounds of subjective evaluation were performed: 1) the first round included submissions from all teams and 2) the second round included several top-scoring teams for further evaluation. The subjective evaluation results of each task were based on the combined results from both rounds. The final ranking will be determined by MOS and released in June.

\begin{table}[t]
\centering
\caption{Results of baseline system on the development simulation set. ``Dev. Simu. Set'' represents development simulation set; ``MA'' represents microphone array; ``Noisy'' represents the speech utterances of development simulation set; ``Enhanced'' represents the enhanced speech utterances by the baseline system.}
\label{table:baseline result}
\footnotesize
\setlength{\tabcolsep}{2.2pt}{
\begin{tabular}{ccccccc}
\hline
Dev. Simu. Set                    & MA                              &          & PESQ  & STOI  & E-STOI & Si-SNR \\ \hline
\multirow{6}{*}{Single MA} & \multirow{2}{*}{Circular}         & Noisy    & 1.514 & 0.824  & 0.693  & 4.566 \\ 
                        &                                    & Enhanced & 1.990 & 0.888  & 0.783  & 9.248 \\ \cline{2-7} 
                        & Linear  & Noisy    & 1.534 & 0.829  & 0.700  & 4.720 \\  
                        & Uniform & Enhanced & 2.035 & 0.893  & 0.790  & 9.445 \\ \cline{2-7} 
                        & Linear  & Noisy    & 1.515 & 0.823  & 0.690  & 4.474 \\  
                        &  Nonuniform  & Enhanced & 1.999 & 0.888  & 0.780  & 9.159 \\ \hline
\multirow{8}{*}{Multiple MA} & \multirow{2}{*}{Circular}         & Noisy    & 1.513 & 0.826  & 0.696  & 4.596 \\  
                        &                                    & Enhanced & 1.997 & 0.890  & 0.785  & 9.271 \\ \cline{2-7} 
                        & Linear & Noisy    & 1.514 & 0.826  & 0.696  & 4.618 \\  
                        & Uniform & Enhanced & 2.007 & 0.891  & 0.786  & 9.260 \\ \cline{2-7} 
                        & Linear & Noisy    & 1.506 & 0.824  & 0.693  & 4.504 \\  
                        & Nonuniform & Enhanced & 1.983 & 0.887  & 0.780  & 9.228 \\ \cline{2-7} 
                        & SNR  & Noisy    & 1.524 & 0.830  & 0.702  & 4.989 \\  
                        &  Selection & Enhanced & 2.023 & 0.893  & 0.788  & 9.526 \\
                        \hline
\end{tabular}
}
\end{table}

\section{Baseline System}

The baseline system we proposed is a complex ratio mask (CRM)~\cite{williamson2015complex} based neural network. The input of the baseline system is the multi-channel speech signals of one MA and the output is the single-channel enhanced speech. After performing Short Time Fourier Transform (STFT), magnitude and phase of input signals will be obtained and then used for estimating the inter-channel phase difference (IPD) of specified microphone pairs:
\begin{equation}
    \text{IPD}_{i,j}=\angle e^{j(\angle\mathbf{O}_{i}-\angle\mathbf{O}_{j})},
\end{equation}
where $\mathbf{O}_{i}$ and $\mathbf{O}_{j}$ denotes the spectrum of observed signal of microphone $i$ and $j$. We concatenate the real and imaginary part of the first microphone's observed signal, as well as the $\cos$IPD of four microphone pairs along the frequency axis to generate $\mathbf{X}\in\mathbb{R}^{6F\times T}$ as the input of the neural network, where $F$ and $T$ denotes the number of frequency bins and frames, respectively. 

A 3-layer real-valued LSTM is used to capture the temporal information of input features. Then a real-valued fully connection (FC) layer is adopted to map the output of LSTM into real and imaginary masks, respectively. CRM is applied to the spectrum of the first microphone channel of observed signal $\mathbf{X}_{0}$ to derive the enhanced speech:
\begin{equation}
    \mathbf{Y}_r=\mathbf{M}_r\mathbf{X}_{0r}-\mathbf{M}_i\mathbf{X}_{0i},
\end{equation}
\begin{equation}
    \mathbf{Y}_i=\mathbf{M}_r\mathbf{X}_{0i}+\mathbf{M}_i\mathbf{X}_{0r},
\end{equation}
where $\mathbf{M}_r$ and $\mathbf{M}_i$ denote the real and imaginary part of CRM, respectively. The setup and code scripts of this baseline system are available on the github page.~\footnote{\url{https://github.com/ConferencingSpeech/ConferencingSpeech2021}}


For training, the provided official training data was used to simulate multi-channel far-field noisy data by convolving single channel signal with RIR. The SNR of simulation training data ranges from 0 to 30 dB. To compress the input feature dimension, for circular MA, 8 channels signals were selected intermittently from all 16 channels signal to calculate the input features, while for linear uniform MA and linear nonuniform MA, first 8 channels signals were selected to calculate the input features. For linear MA, IPDs were calculated among four microphone pairs: (1,5), (2,6), (3,7) and (4,8), while for circular MA, IPDs were calculated among four microphone pairs: (1,9), (3,11), (5,13) and (7,15). The frame length and frame hop of STFT/iSTFT were set to 20ms and 10 ms, respectively. The hidden layer size of the 3-layer LSTM and FC layer were set to 512 and 514, respectively. We trained the model for 18 epochs with Adam optimizer using PyTorch. The initial learning rate was set to 0.001 and would be halved if no improvement on the development set for 2 epochs. The real time factor (RTF) of the baseline system is 0.0425 on Intel Xeon clocked at 2.5 GHz.

For single MA of the development simulation set, three models for circular MA, linear uniform MA and linear nonuniform MA were trained separately, while for multiple MAs, the models trained for single MA were reused and the output with the highest estimated SNR within three models was chosen as the final output. The SNR is calculated by:

\begin{equation}
    \max \limits_{n \in \psi }\{10\text{log}_{10}\frac{\text{RMS}(\textbf{y}_{n})}{\text{RMS}(\textbf{y}_{n}-\textbf{x}_{n})}\},
\end{equation}

where RMS represents root mean square; $\psi$ denotes the all possible microphone array and $\textbf{x}_{n}$ and $\textbf{y}_{n}$ denotes the time domain noisy and enhanced signal of microphone array $n$, respectively.

The results of the baseline system on the development simulation set are shown in Table~\ref{table:baseline result} and they demonstrate that (1) the baseline system effectively enhanced the noisy speech of the development simulation set and (2) the performance of SNR selection in multiple MAs is slightly better than that of single MA model.

\section{Conclusion}\label{Sec:cons}
The ConferencingSpeech challenge is intended to promote far-field multi-channel noise suppression and dereverberation for achieving superior subjective speech quality in real video conferencing scenarios. Two tasks are specially designed for different purposes. Task 1 is on exploring real-time multi-channel speech enhancement methods to achieve excellent perceptual quality and intelligibility of enhanced speech with low latency and no future ``frame'' information. Task 2 is on how to utilize the multiple distributed microphone arrays to improve the performance of speech enhancement. 

Twenty one submissions have been received from seventeen teams. Five teams participated in both tasks. The final ranking, challenge results and related analysis will be announced in June 2021. We believe that this challenge and the published datasets will promote the research and development in far-field multi-channel speech enhancement.

\bibliographystyle{IEEEtran}

\bibliography{mybib}

\end{document}